\documentstyle[12pt]{article}

\newcounter{xpos}
\newcounter{ypos}

\def\Eq{\begin{equation}}
\def\End{\end{equation}}

\def\Eqa{\begin{eqnarray}}
\def\Enda{\end{eqnarray}}

\def\Endl#1{\label{#1} \End}
\def\Endla#1{\label{#1} \Enda}

\def\ord#1{{\cal O}(#1)}

\def\Bp{\begin{picture}}
\def\Ep{\end{picture}}

\def\core#1#2#3#4#5#6{\raisebox{#1pt}{\Bp(#2,#3) \setcounter{xpos}{#4}
  \setcounter{ypos}{#5} #6 \Ep}}

\def\Mr{\addtocounter{xpos}{ 20}}  \def\Ml{\addtocounter{xpos}{-20}}
\def\Mu{\addtocounter{ypos}{ 20}}  \def\Md{\addtocounter{ypos}{-20}}
\def\MR{\addtocounter{xpos}{ 15}}  \def\ML{\addtocounter{xpos}{-15}}
\def\MU{\addtocounter{ypos}{ 15}}  \def\MD{\addtocounter{ypos}{-15}}

\def\Mui{\addtocounter{ypos}{ 24}}  

\def\Cf{\put(\value{xpos},\value{ypos}){\circle*{4}}}   
\def\Cz{\put(\value{xpos},\value{ypos}){\circle{20}}}   
\def\Czi{\put(\value{xpos},\value{ypos}){\circle{18}}   
         \put(\value{xpos},\value{ypos}){\circle{22}}}
\def\Ca{\Cz\Cf}         \def\Cai{\Czi\Cf}               
\def\Cn{\put(\value{xpos},\value{ypos}){\circle{10}}}   

\begin{document}
\begin{titlepage}

\title{{\bf On daisy and superdaisy resummation of the effective
potential at finite temperature}\thanks{Based on talk given at
the {\it 4th Hellenic School on Elementary Particle Physics}, Corfu
(Greece) September 1992}}

\author{{\bf Mariano Quir\'os}\thanks{Work partly supported by CICYT, Spain,
under contract AEN90-0139.} \\
Instituto de Estructura de la Materia, CSIC \\
Serrano 123, 28006-Madrid, Spain  \vspace{1cm} }

\date{}
\maketitle
\begin{abstract}
\noindent
We describe in detail, in the context of the simplest scalar
$\phi^4$ theory, the prescription for resummation
of daisy and superdaisy diagrams in the effective potential
using the solution of the gap equations in the infrared limit.
We find that the latter procedure is consistent provided we
neglect logarithmic terms from the finite-temperature self
energies and from the integration of overlapping momenta.
This amounts to dressing only the zero-mode contribution
to the finite-temperature effective potential.
Improving also the non-zero modes,
would require exactly solving (not in the IR limit) the gap equations.
In general this can only be done in a theory where
all self-energies are momentum independent ({\em e.g.} in the
scalar theory at the symmetric phase $\phi=0$).
However some partial dressing procedures are still
possible in general.
\end{abstract}
\thispagestyle{empty}

\vskip-18.5cm
\noindent
\phantom{bla}
\hfill{{\bf hep-ph/9304284}} \\
\phantom{bla}
\hfill{{\bf IEM-FT-71/93}}

\end{titlepage}

It was recently realized that Sakharov's conditions for
baryogenesis could be accomplished
(in particular that the rate of anomalous $B$-violating
processes is unsuppressed) at high temperatures
in the electroweak theory \cite{KRS}.
This fact revived the interest in the study of the
electroweak phase transition,
mainly because the out-of-equilibrium
condition usually requires the phase
transition to be strongly first-order. The theoretical foundation
for a quantitative treatment of symmetry restoration
and phase transitions at high
temperatures is already twenty years old \cite{KL,W}.
Soon after it was realized \cite{W} that
the perturbative expansion fails at
temperatures close to the critical temperature, because of
infrared (IR) divergences, and it was proposed
to solve the IR problem by the resummation of an infinite set of
the most IR divergent diagrams: {\em i.e.} those belonging to
the daisy and superdaisy classes. Improving the effective
potential in different theories by the inclusion of daisy and
superdaisy diagrams has produced a lot of activity in the field
during the last years \cite{BOOKS}-\cite{BBH2}.
Since there has been some controversy about the correct
resummation procedure concerning the leading infrared divergent
graphs \cite{SHAPO,DINE,ACP,BUCH,AE,BBH2},
I will develop in this note the formalism
which was followed in \cite{EQZ1,EQZ2}
as well as will compare it with other different approaches
recently used by different authors.

We will consider, for simplicity, the theory of a real
scalar field $\phi$, with a tree level potential,
\Eq
V_{{\rm eff}}^{(0)}(\phi)=-\frac{\mu^2}{2}\phi^2+\frac{\lambda}{4}\phi^4
\Endl{treepot}
with positive $\lambda$ and $\mu^2$. At the tree level, the
field-dependent mass of the scalar field is
$m^2(\phi)=3\lambda\phi^2-\mu^2$, and the minimum of
$V_{{\rm eff}}^{(0)}$ corresponds to $v^2=\mu^2/\lambda$, so that
$m^2(v)=2\lambda v^2=2\mu^2$.

At finite temperature, the one-loop effective potential can be
written diagrammatically as\footnote{There is an overall
negative sign in front of all diagrams contributing to the
effective potential and self-energies that
(for simplicity) will be dropped systematically from the figures,
but will be taken into account in the calculation.},
\Eq
V^{(1)}_{{\rm eff}}=\sum_{n=0}^{\infty}V^{(1)}_{[n]}=
\frac{1}{2}\hspace{2pt} \core{-7}{20}{20}{10}{10}{\Ca}
\Endl{one}
where $n$ are the bosonic Matsubara frequencies and
$V^{(1)}_{[n]}$ are the contributions to the one-loop effective
potential from the different frequencies. They can be written to
lowest order in $m/T$ as
\Eq
V^{(1)}_{[0]}=\frac{1}{2} \hspace{2pt}
\core{-7}{20}{20}{10}{10}{\Cz} = -\frac{1}{12\pi}T m^3
\Endl{onezero}
\Eq
\sum_{n=1}^{\infty}V^{(1)}_{[n]}=\frac{1}{2} \hspace{2pt}
\core{-7}{20}{20}{10}{10}{\Cn} =
\frac{1}{24}T^2 m^2+\cdots
\Endl{onen}
where big bubbles denote the contribution from zero modes
and small bubbles the one from all non-zero modes. The
contribution from all modes will be denoted by a big dotted
bubble, {\it i.e.}
\Eq
\core{-7}{20}{20}{10}{10}{\Ca}= \hspace{3pt}
\core{-7}{20}{20}{10}{10}{\Cz} \hspace{5pt}+
\core{-7}{20}{20}{10}{10}{\Cn}
\Endl{decomposition}

For the zero modes ($n=0$) there is a severe infrared problem
in the loop expansion for
values of $\phi$ such that $m(\phi)\ll \lambda T$ at $\vec{p}=0$. At
one-loop the potential (\ref{onezero}) is non-analytic at
$m=0$, while the validity of the perturbative expansion breaks
down at higher-loop order, which contribute powers of $\alpha$
and $\beta$ \cite{W,FENDLEY}
\Eq
\alpha=\lambda \frac{T^2}{m^2},\hspace{.5cm} \beta=\lambda \frac{T}{m}
\Endl{beta}
The usual way out is dressing the zero-modes with daisy and
super-daisy diagrams \cite{W}. This can be done by solving the gap
equations in the IR limit ($n=0,\vec{p} \rightarrow
0$). For the theory defined by Eq.(\ref{treepot}),
and neglecting the terms represented by the ellipsis in
(\ref{onen}), the gap
equation can be diagrammatically written as,

\Eq \Bp(32,3) \put(0,0){\line(1,0){32}} \put(0,2){\line(1,0){32}} \Ep \;=\;
    \Bp(32,1) \put(0,1){\line(1,0){32}} \Ep \;+\;
    \Bp(32,10) \put(0,0){\line(1,0){32}} \put(16,5){\circle{10}} \Ep \;+\;
    \Bp(32,22) \put(0,0){\line(1,0){32}}
      \put(16,12){\circle{18}\circle{22}} \Ep \;+\;
    \raisebox{-11pt}{\Bp(44,22) \put(0,11){\line(1,0){10}}
      \put(22,11){\circle{18}\circle{22}}
      \put(34,11){\line(1,0){10}} \Ep}
\Endl{gap}
where a double line represents a dressed zero-mode propagator.
Using the approximation in Eq.(\ref{onen}) the self-energies can
be written as
\Eq
   \Bp(32,10) \put(0,0){\line(1,0){32}} \put(16,5){\circle{10}} \Ep
\;=\frac{\lambda}{4}T^2 \;+\; \cdots
\Endl{an}
\Eq
 \Bp(32,22) \put(0,0){\line(1,0){32}}
      \put(16,12){\circle{22}} \Ep \;
=-\frac{3\lambda T m}{4\pi}
\Endl{azero}
\Eq
 \raisebox{-10pt}{\Bp(32,11) \put(0,11){\line(1,0){10}}
      \put(16,11){\circle{11}}
      \put(22,11){\line(1,0){10}} \Ep}  \equiv 0\;+\; \cdots
\Endl{bn}
\Eq
 \raisebox{-11pt}{\Bp(44,22) \put(0,11){\line(1,0){10}}
      \put(22,11){\circle{22}}
      \put(34,11){\line(1,0){10}} \Ep}  =
-\frac{9\lambda^2 \phi^2 T}{4\pi m}
\Endl{bzero}
and the gap equation (\ref{gap}) as
\Eq
{\displaystyle
M^2=m^2+\frac{\lambda T^2}{4}-\frac{3\lambda T M}{4\pi}
-\frac{9\lambda^2 \phi^2 T}{4\pi M}  }
\Endl{gapeq}
where $M$ is the solution to (\ref{gapeq}). In the approximation
of Eq.(\ref{onen}) the small bubbles are constant proportional
to $T^2$ (\ref{an}) or zero (\ref{bn})
(the ellipsis is neglected), and so they do not have
to be dressed. Going beyond this approximation, also small
bubbles would need to be dressed. However for bubbles of the kind
(\ref{bn}) and (\ref{bzero}) the IR limit
would no longer be justified at all.

Now we will see how the daisy and superdaisy diagrams amount to a
resummation in the loop expansion of the effective potential
which can therefore be written in terms of the solution to the gap
equation (\ref{gapeq}). In the order of approximation we are working
only the zero modes need to be dressed, and only $V^{(1)}_{[0]}$
in Eq.(\ref{onezero}) is improved, while
$\sum_{n \neq 0}V^{(1)}_{[n]}$ in Eq.(\ref{onen})
does not have any IR problem and can be considered as a good estimate.
We will prove the resummation to four-loop order though also functional
methods \cite{CJT} can be used \cite{ACP}.

The loop expansion of the effective potential will be written as
\Eq
V_{{\rm eff}}=\sum_{\ell =0}^{\infty} V_{{\rm eff}}^{(\ell)}
\;+\; {\rm non-(super)daisies}
\Endl{loopexp}
where $V_{{\rm eff}}^{(\ell)}$ indicates the contribution to the effective
potential from $\ell$-loop daisy and superdaisy
diagrams. Non-(super)daisy diagrams contribute to the effective
potential to $\ord{\beta^2}$ \cite{FENDLEY,EQZ1}. At least, to $\ord{\beta}$
it is consistent to keep only diagrams of daisy and superdaisy classes.
$V_{{\rm eff}}^{(1)}$ was given in Eq.(\ref{one}), while
$V_{{\rm eff}}^{(2)}$ and $V_{{\rm eff}}^{(3)}$ can be written as

\Eq
V_{{\rm eff}}^{(2)}=\frac{1}{8} \hspace{5pt}
\core{-17}{20}{40}{10}{10}{\Ca\Mu\Ca}+\frac{1}{12}
\hspace{5pt}\raisebox{-7pt}{\Bp(20,20)\setcounter{xpos}{10}
\setcounter{ypos}{10} \Cz \put(10,5){\circle*{3}}
\put(10,15){\circle*{3}}
\put(0,10){\line(1,0){20}}
\Ep}
\Endl{two}
\vspace{.5cm}
\Eq
V_{{\rm eff}}^{(3)}=\frac{1}{16} \hspace{5pt}
\core{-25}{20}{60}{10}{10}{\Ca\Mu\Ca\Mu\Ca}+\frac{1}{8}
\hspace{5pt}\raisebox{-17pt}{\Bp(20,40)\setcounter{xpos}{10}
\setcounter{ypos}{10} \Cz\Mu\Ca \put(10,5){\circle*{3}}
\put(10,15){\circle*{3}}
\put(0,10){\line(1,0){20}}
\Ep}+\frac{1}{16}
\hspace{5pt}\raisebox{-7pt}{\Bp(20,20)\setcounter{xpos}{10}
\setcounter{ypos}{10} \Cz \put(10,5){\circle*{2}}
\put(10,15){\circle*{2}} \put(10,10){\circle*{2}}
\put(0,12.5){\line(1,0){20}}
\put(0,7.5){\line(1,0){20}}
\Ep}
\Endl{three}
where we are putting dots everywhere to remember that all modes
(zero and non-zero modes) are contributing in the loop propagators,
and the numerical pre-factors in front of (\ref{two})
and (\ref{three}) are the symmetry factors of the corresponding
diagrams.

Using the approximation in (\ref{an}), (\ref{bn}), we can rearrange
the loop expansion in (\ref{one}), (\ref{two}) and (\ref{three}) as

\Eq
V_{{\rm eff}}^{(\ell)}=V_{{\rm daisy}}^{(\ell)}+
V_{{\rm superdaisy}}^{(\ell)}
+\Delta^{(\ell)}
\left[
-\frac{1}{8} \hspace{5pt} \core{-17}{20}{40}{10}{10}{\Czi\Mui\Czi}
-\frac{1}{6} \hspace{5pt} \core{-7}{20}{20}{10}{10}{\Czi
\put(0,9){\line(1,0){20}} \put(0,11){\line(1,0){20}}}
\hspace{10pt} \right]
\Endl{rearrangement}
where $\Delta^{(\ell)}[\cdots]$ means the contribution to $[\cdots]$ from
$\ell$-loop diagrams and the double line is given by Eq.(\ref{gap}).
This decomposition is well defined for $\ell \geq 2$.
Next we give the results for two and three-loop diagrams.

\vspace{.5cm}
{\bf Two-loop}
\Eq
V_{{\rm daisy}}^{(2)}=\hspace{2pt} \frac{1}{4}\hspace{5pt}
\core{-12}{20}{40}{10}{10}{\Cz\MU\Cn}
+\hspace{2pt}\frac{1}{4} \hspace{5pt}
\core{-17}{20}{40}{10}{10}{\Cz\Mu\Cz}
+\hspace{2pt}\frac{1}{4}
\hspace{5pt}\raisebox{-7pt}{\Bp(20,20)\setcounter{xpos}{10}
\setcounter{ypos}{10} \Cz
\put(0,10){\line(1,0){20}}
\Ep}
\Endl{twodaisy}

\Eq
V_{{\rm superdaisy}}^{(2)}\;=\;0
\Endl{twosuper}
\Eq
\Delta^{(2)}=
-\hspace{2pt}\frac{1}{8} \hspace{5pt}
\core{-17}{20}{40}{10}{10}{\Cz\Mu\Cz}
-\hspace{2pt}\frac{1}{6}
\hspace{5pt}\raisebox{-7pt}{\Bp(20,20)\setcounter{xpos}{10}
\setcounter{ypos}{10} \Cz
\put(0,10){\line(1,0){20}}
\Ep}
\Endl{twodelta}

\vspace{.5cm}
{\bf Three-loop}
\Eq
V_{{\rm daisy}}^{(3)}=\frac{1}{16} \hspace{5pt}
\core{-17}{20}{40}{10}{20}{\Cz\MU\Cn\MD\MD\Cn}
+\frac{1}{8} \hspace{5pt}
\core{-22}{20}{50}{10}{25}{\Cz\MU\Cn\MD\Md\Cz}
+\frac{1}{16} \hspace{5pt}
\core{-27}{20}{60}{10}{30}{\Cz\Mu\Cz\Md\Md\Cz}
+\frac{1}{8}
\hspace{5pt}\raisebox{-12pt}{\Bp(20,30)\setcounter{xpos}{10}
\setcounter{ypos}{10} \Cz\MU\Cn
\put(0,10){\line(1,0){20}}
\Ep}
+\frac{1}{8}
\hspace{5pt}\raisebox{-12pt}{\Bp(20,40)\setcounter{xpos}{10}
\setcounter{ypos}{10} \Cz\Mu\Cz
\put(0,10){\line(1,0){20}}
\Ep}
+\frac{1}{16}
\hspace{5pt}\raisebox{-7pt}{\Bp(20,20)\setcounter{xpos}{10}
\setcounter{ypos}{10} \Cz
\put(0,12.5){\line(1,0){20}}
\put(0,7.5){\line(1,0){20}}
\Ep}
\Endl{threedaisy}

\Eq
V_{{\rm superdaisy}}^{(3)}=
\frac{1}{8} \hspace{5pt}
\core{-4}{20}{50}{10}{25}{\Cz\MU\Cn\MD\Md\Cz}
+\frac{1}{8} \hspace{5pt}
\core{-7}{20}{60}{10}{30}{\Cz\Mu\Cz\Md\Md\Cz}
+\frac{1}{4}
\hspace{5pt}\raisebox{-7pt}{\Bp(20,30)\setcounter{xpos}{10}
\setcounter{ypos}{10} \Cz\MU\Cn
\put(0,10){\line(1,0){20}}
\Ep}
+\frac{3}{8}
\hspace{5pt}\raisebox{-7pt}{\Bp(20,40)\setcounter{xpos}{10}
\setcounter{ypos}{10} \Cz\Mu\Cz
\put(0,10){\line(1,0){20}}
\Ep}
+\frac{1}{4}
\hspace{5pt}\raisebox{-7pt}{\Bp(20,20)\setcounter{xpos}{10}
\setcounter{ypos}{10} \Cz
\put(0,12.5){\line(1,0){20}}
\put(0,7.5){\line(1,0){20}}
\Ep}
\Endl{threesuper}

\Eq
\Delta^{(3)}=
-\frac{1}{8} \hspace{5pt}
\core{-12}{20}{50}{10}{25}{\Cz\MU\Cn\MD\Md\Cz}
-\frac{1}{8} \hspace{5pt}
\core{-17}{20}{60}{10}{30}{\Cz\Mu\Cz\Md\Md\Cz}
-\frac{1}{4}
\hspace{5pt}\raisebox{-12pt}{\Bp(20,30)\setcounter{xpos}{10}
\setcounter{ypos}{10} \Cz\MU\Cn
\put(0,10){\line(1,0){20}}
\Ep}
-\frac{3}{8}
\hspace{5pt}\raisebox{-17pt}{\Bp(20,40)\setcounter{xpos}{10}
\setcounter{ypos}{10} \Cz\Mu\Cz
\put(0,10){\line(1,0){20}}
\Ep}
-\frac{1}{4}
\hspace{5pt}\raisebox{-7pt}{\Bp(20,20)\setcounter{xpos}{10}
\setcounter{ypos}{10} \Cz
\put(0,12.5){\line(1,0){20}}
\put(0,7.5){\line(1,0){20}}
\Ep}
\Endl{threedelta}

We can see from (\ref{two}) and (\ref{twodaisy}) that the symmetry factors
for $\ell=2$ do not match the combinatorics for daisy resummation.
However including (\ref{twodelta}) the matching is accomplished
as can be seen from the coefficients in (\ref{two}) and the last two terms
in (\ref{twodaisy}) and (\ref{twodelta})
\Eqa
\frac{1}{4}-\frac{1}{8}=&{\displaystyle \frac{1}{8}} \nonumber \\
\frac{1}{4}-\frac{1}{6}=&{\displaystyle  \frac{1}{12}}
\Endla{twocheck}
On the other hand, we have seen from
(\ref{threesuper},\ref{threedelta}) that
$V_{{\rm superdaisy}}^{(3)}
+\Delta^{(3)}=0$. The reason being that all the
diagrams in (\ref{three})
could be interpreted either as daisies or as
superdaisies. Therefore we have considered all of them
as daisies, because their coefficients match the correct combinatorics for
resummation. This can be seen by comparison with the corresponding
coefficients in (\ref{threedaisy}) and it is a general feature of daisy
diagrams for $\ell \geq 3$.
However, for $\ell \geq 4$ there are diagrams that can
never be considered as daisies. In that case the previous cancellation
does not hold, but still the equation (\ref{rearrangement}) is satisfied.
As an example we will consider the theory at the origin ({\em i.e.} at
$\phi=0$) to four-loop order.

\vspace{.5cm}
{\bf Four-loop}

The contributions to (\ref{loopexp}) and (\ref{rearrangement})
can be written as

\Eq
V^{(4)}_{{\rm eff}}(0)=
\frac{1}{48} \hspace{5pt}
\core{-17}{60}{40}{30}{20}{\Ca\Mr\Ca\Ml\Mu\Ca\Md\Ml\Ca}
+\frac{1}{32} \hspace{5pt}
\core{-35}{20}{80}{10}{10}{\Ca\Mu\Ca\Mu\Ca\Mu\Ca}
\Endl{four}
\vspace{0.5cm}
\Eq
V_{{\rm daisy}}^{(4)}(0)=
\frac{1}{48} \hspace{5pt}
\core{-7}{40}{30}{20}{10}{\Cz\MU\Cn\MD\MR\Cn\ML\ML\Cn}
+\frac{3}{48} \hspace{5pt}
\core{-7}{50}{30}{30}{10}{\Cz\Ml\Cz\Mr\MU\Cn\MD\MR\Cn}
+\frac{3}{48} \hspace{5pt}
\core{-7}{50}{40}{30}{10}{\Cz\Ml\Cz\Mr\Mu\Cz\Md\MR\Cn}
+\frac{1}{48} \hspace{5pt}
\core{-17}{60}{40}{30}{20}{\Cz\Mr\Cz\Ml\Mu\Cz\Md\Ml\Cz}
\Endl{fourdaisy}

\Eq
V_{{\rm superdaisy}}^{(4)}(0)=
\frac{1}{16} \hspace{5pt}
\core{-7}{30}{50}{10}{10}{\Cz\Mu\Cz\MR\Cn\ML\MU\Cn}
+\frac{1}{8} \hspace{5pt}
\core{-7}{30}{60}{10}{10}{\Cz\Mu\Cz\MR\Cn\ML\Mu\Cz}
+\frac{1}{16} \hspace{5pt}
\core{-7}{40}{60}{10}{10}{\Cz\Mu\Cz\Mr\Cz\Ml\Mu\Cz}
+\frac{1}{16} \hspace{5pt}
\core{-10}{20}{60}{10}{20}{\Cz\MD\Cn\MU\Mu\Cz\MU\Cn}
+\frac{3}{16} \hspace{5pt}
\core{-22}{20}{70}{10}{5}{\Cn\MU\Cz\Mu\Cz\Mu\Cz}
+\frac{2}{16} \hspace{5pt}
\core{-22}{20}{80}{10}{10}{\Cz\Mu\Cz\Mu\Cz\Mu\Cz}
\Endl{foursuper}

\Eq
\Delta^{(4)}(0)=
-\frac{1}{16} \hspace{5pt}
\core{-7}{30}{50}{10}{10}{\Cz\Mu\Cz\MR\Cn\ML\MU\Cn}
-\frac{1}{8} \hspace{5pt}
\core{-7}{30}{60}{10}{10}{\Cz\Mu\Cz\MR\Cn\ML\Mu\Cz}
-\frac{1}{16} \hspace{5pt}
\core{-7}{40}{60}{10}{10}{\Cz\Mu\Cz\Mr\Cz\Ml\Mu\Cz}
-\frac{1}{32} \hspace{5pt}
\core{-10}{20}{60}{10}{20}{\Cz\MD\Cn\MU\Mu\Cz\MU\Cn}
-\frac{1}{8} \hspace{5pt}
\core{-22}{20}{70}{10}{5}{\Cn\MU\Cz\Mu\Cz\Mu\Cz}
-\frac{3}{32} \hspace{5pt}
\core{-22}{20}{80}{10}{10}{\Cz\Mu\Cz\Mu\Cz\Mu\Cz}
\Endl{fourdelta}

The first diagram in (\ref{four}) can be (and it is) considered
as a daisy diagram in (\ref{fourdaisy}). For that reason the
coefficients of the first three terms in (\ref{foursuper}) and
(\ref{fourdelta}) are equal and opposite in sign. The second diagram in
(\ref{four}) can {\it never} be considered as a daisy diagram. So it
contributes to the last three terms of
(\ref{foursuper}) and (\ref{fourdelta}) in such a
way that their coefficients match to those of the second term of
(\ref{four}). In particular

\Eqa
\frac{1}{16}-\frac{1}{32}=& {\displaystyle \frac{1}{32}} \nonumber \\
\frac{3}{16}-\frac{1}{8}=&{\displaystyle 2\times \frac{1}{32}} \\
\frac{2}{16}-\frac{3}{32}=& {\displaystyle \frac{1}{32}} \nonumber
\Endla{check}
for the last three terms, respectively.

\vspace{.5cm}
{\bf All-loop}

Summarizing the above results, we can write the final equation:

\Eq
V_{{\rm eff}}=\frac{1}{2}\core{-7}{20}{20}{10}{10}{\Cn}
+\frac{1}{2}\hspace{5pt} \core{-7}{20}{20}{10}{10}{\Czi} \hspace{5pt}
-\frac{1}{8} \hspace{5pt} \core{-17}{20}{40}{10}{10}{\Czi\Mui\Czi}
-\frac{1}{6} \hspace{5pt} \core{-7}{20}{20}{10}{10}{\Czi
\put(0,9){\line(1,0){20}} \put(0,11){\line(1,0){20}}}
\Endl{final}
where the double line indicates the solution of the gap equation
(\ref{gap}) and (\ref{gapeq}), in the approximation of
Eqs.(\ref{an}) and (\ref{bn}). Using the explicit solution to
(\ref{gapeq}) we can write (\ref{final}) as
\Eq
V_{{\rm eff}}=-\frac{\mu^2}{2}\phi^2+\frac{\lambda}{4}\phi^4
+\frac{1}{24}T^2 m^2 -\frac{1}{12\pi}T M^3+\cdots
-\frac{3\lambda}{64 \pi^2}T^2 M^2
-\frac{6}{32\pi^2}\lambda^2 \phi^2 T^2
\Endl{potential}
which agrees with the result of Amelino-Camelia and Pi
\cite{ACP}, who used functional methods \cite{CJT}
and computed (\ref{gap}) and (\ref{gapeq})
to zeroth order in $\gamma$
\Eq {\displaystyle
\gamma \equiv \frac{\phi^2}{T^2}  }
\Endl{gamma}
For that reason the last term in (\ref{potential}) was missing
from Eq.(3.25) in Ref. \cite{ACP}. The others agree. However we
will see that, though numerically unimportant in the scalar case
because $\gamma \ll 1$, all terms in (\ref{gapeq}) should be
considered in the $\beta$-expansion.

In the improved theory of zero-modes defined by (\ref{gapeq})
and (\ref{potential}), the expansion parameters $\alpha$ and
$\beta$ in (\ref{beta}) become\footnote{We will keep
for notational simplicity the
same names $\alpha$ and $\beta$ for the expansion parameters in
the improved (as in the unimproved) theory.}
\Eq {\displaystyle
\alpha=\lambda \frac{T^2}{{\cal M}^2} \sim 1  }
\Endl{alphaimp}
which is summed to all orders, and
\Eq {\displaystyle
\beta=\lambda \frac{T}{{\cal M}} \sim \lambda^{\frac{1}{2}}  }
\Endl{betaimp}
which remains as the only expansion parameter, where ${\cal M}$
is the Debye mass
\Eq {\displaystyle
{\cal M}^2=m^2(\phi)+\frac{\lambda}{4} T^2  }
\Endl{debye}

By expanding the solution of Eq.(\ref{gapeq}) to different
orders in $\beta$ we can obtain the effective potential
(\ref{potential}) to the corresponding order of approximation.
To illustrate the procedure we will first obtain the solution to
$\ord{\beta^0}$. In that case we have
\Eq
M^2={\cal M}^2
\Endl{gapzero}
and the effective potential is given by
\Eq
V_{{\rm eff}}=-\frac{\mu^2}{2}\phi^2+\frac{\lambda}{4}\phi^4
+\frac{1}{24}T^2 m^2 -\frac{1}{12\pi}T {\cal M}^3+\cdots
\Endl{potentialzero}
This approximation has been worked out in \cite{CARRINGTON}.
The last two terms in (\ref{final}) and (\ref{potential}) do not
contribute to this order\footnote{In the language of Ref.
\cite{EQZ2} there is no {\em combinatorial} term to this order.}
since they start to $\ord{\beta}$.
The solution to $\ord{\beta}$ is equally easy to be worked out.
{}From (\ref{gapeq}) one can write
\Eq {\displaystyle
M^2={\cal M}^2-\frac{3\lambda}{4\pi}{\cal M}T
-\frac{9\lambda^2}{4\pi}\phi^2 \frac{T}{{\cal M}}   }
\Endl{gapbeta}
where the first term is the leading order result,
Eq.(\ref{gapzero}), the second
term is $\ord{\beta}$ and the third term is
$\ord{\alpha\beta\gamma}$. As we said above, the last term
cannot be neglected in the $\beta$-expansion, though it can be
numerically unimportant. Replacing (\ref{gapbeta}) in
(\ref{potential}) and expanding again to $\ord{\beta}$ we can
obtain the corresponding approximation to the effective
potential, given by
\Eq
V_{{\rm eff}}=-\frac{\mu^2}{2}\phi^2+\frac{\lambda}{4}\phi^4
+\frac{1}{24}T^2 m^2 -\frac{1}{12\pi}T {\cal M}^3+\cdots
+\frac{(6-3)}{64 \pi^2}\lambda T^2 {\cal M}^2
+\frac{(9-6)}{32\pi^2}\lambda^2 \phi^2 T^2
\Endl{potentialbeta}
This solution was presented in \cite{EQZ2}. We can easily check that
the last two terms in (\ref{potentialbeta}) are $\ord{\beta}$
corrections to the fourth term. They come partly from the
expansion (\ref{gapbeta}) and partly from the last two terms in
(\ref{final}).

We have already compared our results with those of Ref.
\cite{ACP}. There is another, recent, proposal by P. Arnold and
O. Espinosa \cite{AE} who have advocated a hybrid method using a
partial resummation of the leading contribution of
$n\neq 0$ bubbles followed by an ordinary
loop expansion \cite{P}. One can define a {\it partially dressed}
propagator as
\Eq
 \Bp(40,3) \put(20,2){\circle*{3}} \put(0,2){\line(1,0){40}} \Ep \;=\;
 \Bp(40,3) \put(0,2){\line(1,0){40}} \Ep \;+\;
 \Bp(40,10) \put(0,2){\line(1,0){40}} \put(20,4.5){\circle{5}} \Ep \;+\;
 \Bp(40,10) \put(0,2){\line(1,0){40}} \put(10,4.5){\circle{5}}
 \put(30,4.5){\circle{5}} \Ep \;+\;
\Bp(40,10) \put(0,2){\line(1,0){40}} \put(10,4.5){\circle{5}}
\put(20,4.5){\circle{5}} \put(30,4.5){\circle{5}} \Ep
 +\cdots
\Endl{partgap}
where the tiny bubble propagator is defined as,
\Eq
\Bp(30,10) \put(0,0){\line(1,0){30}} \put(15,2.5){\circle{5}} \Ep
\;=\;\frac{\lambda}{4} T^2 \;,
\Endl{partan}
and a {\it partially resummed} loop expansion as
\Eq
V_{{\rm eff}}=V^{(0)}_{{\rm eff}}+
\frac{1}{2}\hspace{2pt} \core{-7}{20}{20}{10}{10}{\Ca
\put(10,20){\circle*{3}} }+
\frac{1}{8} \hspace{5pt}
\core{-17}{20}{40}{10}{10}{\Ca\Mu\Ca \put(10,0){\circle*{3}}
\put(10,40){\circle*{3}} }+\frac{1}{12}
\hspace{5pt}\raisebox{-12pt}{\Bp(28,28)\setcounter{xpos}{14}
\setcounter{ypos}{14} \put(14,14){\circle{28}}
\put(14,7){\circle*{4}}
\put(14,22){\circle*{4}}
\put(14,0){\circle*{3}}
\put(14,14){\circle*{3}}
\put(14,28){\circle*{3}}
\put(0,14){\line(1,0){28}}
\Ep} +\cdots
\Endl{partloop}
It is easy to check that using the approximation of
Eqs.(\ref{an})-(\ref{bzero}), and ignoring overlapping momenta,
one recovers, to $\ord{\beta}$, the same result as that of
Eq.(\ref{potentialbeta}).

Other authors \cite{DINE,BBH1} have proposed computing tadpoles,
instead of vacuum diagrams, to exhibit some features of the
improved theory, {\it e.g.} resummation properties and the
absence of a linear term in $m(\phi)$ in the final effective
potential. Since the tadpole is nothing else than the
$\phi$-derivative of the effective potential, there can be no
difference between both formalisms. In fact, by comparison between the
contents of this note and those in \cite{BBH2} one can
easily see that the resummation properties of the tadpole diagrams
are inherited from the corresponding ones in vacuum diagrams.
However, in our
opinion, the tadpole formalism has two practical drawbacks:
i) The classification of tadpole diagrams is much more
involved than the classification of vacuum diagrams; ii) To
obtain the effective potential the tadpole has to be integrated,
which can be a non-trivial operation since it depends on the
solution of the gap equation.
However, at the end, both methods should yield the same result.

In our approximation of Eqs.(\ref{onen}), (\ref{an}) and
(\ref{bn}) we have been neglecting all logarithmic terms, which
amounted to not dressing non-zero modes. Including them would
amount to write the previous equations as
\Eq
\frac{1}{2} \hspace{2pt}
\core{-7}{20}{20}{10}{10}{\Cn} =
\frac{1}{24}T^2 m^2
-\frac{m^4}{64\pi^2}\log \frac{Q^2}{c_B T^2}
+\cdots
\Endl{onenlog}
where $Q$ is the renormalization scale in the $\overline{{\rm
DR}}$ scheme and $\log(c_B)=3.9076$,
\Eq
   \Bp(32,10) \put(0,0){\line(1,0){32}} \put(16,5){\circle{10}} \Ep
\;=\frac{\lambda}{4}T^2
-\frac{3 \lambda m^2}{16 \pi^2}\log \frac{Q^2}{c_B T^2}
+\cdots
\Endl{anlog}
\Eq
 \raisebox{-10pt}{\Bp(32,11) \put(0,11){\line(1,0){10}}
      \put(16,11){\circle{11}}
      \put(22,11){\line(1,0){10}} \Ep} =
-\frac{9 \lambda^2 \phi^2}{8 \pi^2}\log \frac{Q^2}{c_B T^2}
+\cdots
\Endl{bnlog}
In that case the gap equation should be written as
\Eq \Bp(32,3) \put(0,0){\line(1,0){32}} \put(0,2){\line(1,0){32}} \Ep \;=\;
    \Bp(32,1) \put(0,1){\line(1,0){32}} \Ep \;+\;
    \Bp(32,22) \put(0,0){\line(1,0){32}}
      \put(16,12){\circle{18}\circle{22}\circle*{4}} \Ep \;+\;
    \raisebox{-11pt}{\Bp(44,22) \put(0,11){\line(1,0){10}}
      \put(22,11){\circle{18}\circle{22}\circle*{4}}
      \put(34,11){\line(1,0){10}} \Ep}
\Endl{gaplog}
and other diagrams should be added to those in
(\ref{twodaisy})-(\ref{fourdelta}). In particular small bubbles
dressed by insertions of self-energies of
the kind (\ref{bzero}). But the latter
are computed in the IR limit, which is not justified at all if
the external mode has $n\neq 0$. If we insist in keeping the
logarithmic terms (which in principle are expected to constitute
small corrections to the leading
contribution) we should give up resummation in the sense of the gap
equation (\ref{gaplog}). A possibility is making a $\gamma$
expansion for the gap equation, and defining an $\ord{\gamma^0}$
gap equation as
\Eq \Bp(32,3) \put(0,0){\line(1,0){32}} \put(0,2){\line(1,0){32}}
\put(16,1){\circle*{4}} \Ep \;=\;
    \Bp(32,1) \put(0,1){\line(1,0){32}} \Ep \;+\;
    \Bp(32,23) \put(0,0){\line(1,0){32}}
      \put(16,12){\circle{18}\circle{22}\circle*{4}}
      \put(16,23){\circle*{4}} \Ep
\Endl{gappart}
where the dotted double line means the solution of the truncated
gap equation\footnote{Of course the gap equation (\ref{gappart}),
and (\ref{gapeqpart}), is exact at the origin $\phi=0$, where
$\gamma \equiv 0$.}, {\it i.e.}
\Eq
{\displaystyle
M^2=m^2+\frac{\lambda T^2}{4}-\frac{3\lambda T M}{4\pi}
-\frac{3 \lambda M^2}{16 \pi^2}\log \frac{Q^2}{c_B T^2}
+\cdots }
\Endl{gapeqpart}
However, we should pay attention to the fact that the last term
in the full gap equation (\ref{gapeq}) is
$\ord{\alpha\beta\gamma}$ and solving it, and giving the
improved effective potential to some order in $\beta$ implies
that we should consider the same order in $\gamma$ by adding
loop diagrams. In this way repeating the whole above procedure
we would find that to $\ord{\beta}$ one can write the effective
potential as
\Eq
V_{{\rm eff}}=
\frac{1}{2}\hspace{5pt} \core{-7}{22}{22}{11}{11}{\Cai
\put(11,0){\circle*{4}}}
-\frac{1}{8} \hspace{5pt} \core{-17}{22}{47}{11}{11}{\Cai\Mui\Cai
\put(11,0){\circle*{4}} \put(11,46){\circle*{4}}}
+\frac{1}{12} \hspace{5pt} \core{-7}{22}{22}{11}{11}{\Cai
\put(1,10){\line(1,0){20}} \put(1,12){\line(1,0){20}}
\put(11,0){\circle*{4}}  \put(11,5.5){\circle*{4}}
\put(11,16.5){\circle*{4}}  \put(11,22){\circle*{4}} }
+\ord{\beta^2}
\Endl{finalpart}
where the equation (\ref{gapeqpart}) is solved to $\ord{\beta}$.
This solution coincides at $\phi=0$ with that in Ref. \cite{ACP}
and for all values of $\phi$ with that in Ref. \cite{AE}.
One can check that the logarithmic corrections which appear are
both from the logarithms in (\ref{onenlog}), (\ref{anlog}),
(\ref{bnlog}), and from the overlapping momenta whose integral
is explicitly considered. To
higher order in $\beta$ more terms should be added to
(\ref{finalpart}) but care should be taken not to commit
overcounting and non-(super)daisy diagrams should be considered.

In conclusion we have reviewed the prescription for resummation
of daisy and superdaisy diagrams in the effective potential
using the solution of the gap equations in the IR limit. We have
found that the latter procedure is consistent provided we
neglect logarithmic terms in the finite-temperature self
energies. This amounts to improving only the zero-mode
finite temperature contribution
to the potential. Trying to go beyond this approximation, and
improving also the non-zero modes,
would be inconsistent with the usual daisy
and superdaisy resummation, based on the solution of the gap
equations in the IR limit.
It would require exact (not in the IR limit) calculation of self-energy
insertions and exact solutions of the gap equations.
This can only be done in a theory where
all self-energies are momentum independent ({\em e.g.} in a
scalar theory at the origin $\phi=0$).
The corrections due to the dressing
of non-zero modes are proportional to logarithmic terms arising
from the self-energies and from integration of overlapping
momenta in calculation of diagrams.
In a theory with fermions, the corrections due to dressing of
bosonic non-zero modes should be similar to corrections due to
fermion dressing which presumably should also be considered.
If we insist in keeping
these corrections we should give up the usual daisy and superdaisy
resummation. Some partial dressing procedures are
still possible.

\vspace{.5cm}
{\Large{\bf Acknowledgements}}
\vspace{.5cm}

The contents of this note are based on joint work done in
collaboration with Jose Ramon Espinosa and Fabio Zwirner to whom
I want to express my deep gratitude. I also thank G. Boyd, D. Brahm,
A. Linde and M.E. Shaposhnikov for useful (and sometimes endless)
discussions.

\end{document}